# Ghost anti-crossings caused by interlayer umklapp hybridization of bands in 2D heterostructures

*Abigail J. Graham¹, Johanna Zultak²,³, Matthew J. Hamer²,³, Viktor Zolyomi²,³, Samuel Magorrian²,³, Alexei Barinov⁴, Viktor Kandyba⁴, Alessio Giampietri⁴, Andrea Locatelli⁴, Francesca Genuzio⁴, Natalie C. Teutsch¹, Cuauhtémoc Salazar¹, Nicholas D. M. Hine¹, Vladimir I. Fal'ko\*²,³,⁵, Roman V. Gorbachev\*²,³,⁵, Neil R. Wilson\*¹*

¹Department of Physics, University of Warwick, Coventry, CV4 7AL, U.K.

²National Graphene Institute, University of Manchester, Booth St East, Manchester, M13 9PL, U.K.

³School of Physics and Astronomy, University of Manchester, Oxford Road, Manchester, M13 9PL, U.K.

⁴Elettra – Sincrotrone Trieste, S.C.p.A, Basovizza (TS), 34149, Italy

⁵Henry Royce Institute, Oxford Road, Manchester, M13 9PL, U.K.

*Email: Vladimir.Falko@manchester.ac.uk.

*Email: Roman@manchester.ac.uk.

*Email: Neil.Wilson@warwick.ac.uk.

## Abstract

In two-dimensional heterostructures, crystalline atomic layers with differing lattice parameters can stack directly one on another. The resultant close proximity of atomic lattices with differing periodicity can lead to new phenomena. For umklapp processes, this opens the possibility for interlayer umklapp scattering, where interactions are mediated by the transfer of momenta to or from the lattice in the neighbouring layer. Using angle-resolved photoemission spectroscopy to study a graphene on InSe heterostructure, we present evidence that interlayer umklapp processes can cause hybridization between bands from neighbouring layers in regions of the Brillouin zone where bands from only one layer are expected, despite no evidence for Moiré-induced replica bands. This phenomenon manifests itself as "ghost" anti-crossings in the InSe electronic dispersion. Applied to a range of suitable two-dimensional material pairs, this phenomenon of interlayer umklapp hybridization can be used to create strong mixing of their electronic states, giving a new tool for twist-controlled band structure engineering.

## Introduction

Crystalline periodicity modifies the interpretation of the momentum conservation law for electronic and optical processes in solids. It gives rise to a periodicity of the electronic dispersion in momentum space, so that, according to Bloch's theorem [1], the bandstructure is uniquely defined within one (the first) Brillouin zone. All processes in a crystal can then be divided into two types: those with small momentum differences that can be described within the first Brillouin zone and those where a large momentum transfer requires the involvement of other Brillouin zones. In the latter case, a momentum transfer $\hbar G$ to the crystalline lattice, where $G$ is one of the reciprocal lattice vectors, satisfies the conservation of momentum and was dubbed *Umklapprozesse* (Umklapp processes) by Peierls [2]. When applied to heterostructures of two-dimensional materials (2DM), umklapp scattering from Moiré superlatttices has been shown to open new channels for electron kinetics [3] and optical transitions [4].

2DMs represent a broad class of compounds where atomic planes formed by strong in-plane covalent bonding are held together by a weak van der Waals interaction. These weak out-of-plane forces enable the stacked assembly of 2DM heterostructures (2DHS), where consecutive layers may involve atomic planes of different compounds with arbitrary lattice constants and orientation, with atomically clean interfaces [5–7] which allow neighbouring layers in the heterostructure to influence each other, in particular through tunneling. Tunneling across clean interfaces is subject to momentum conservation [8,9], so that it is resonantly enhanced in the part of momentum space where the bands of two 2DM intersect, causing resonant interlayer hybridization. Dramatic bandstructure modifications through resonant interlayer hybridization have been studied in twisted bilayers of graphene [10–12], graphene on single-crystal metal substrates [13,14], and graphene with other 2DM [15,16], leading to band anti-crossings and, potentially, to van Hove singularities in the density of states. Here, we demonstrate that interlayer umklapp processes in resonant tunneling lead to the appearance of additional features in the hybridized band structures of 2DHS.

## Results and Discussion

An example of such an effect is illustrated in Figure 1, where angle-resolved photoemission spectroscopy with submicrometre spatial resolution (µARPES) has been used to probe the valence band structure in a graphene on InSe 2DHS. In Figure 1b we sketch the valence band dispersion of monolayer InSe (unfolded over the second Brillouin zone replicas marked by $\mu = 0,1\ldots5$) and the $\pi$-bands of graphene. Notably, no interlayer band crossing occurs in the first Brillouin zone of InSe so one would not expect any resonant hybridization of electronic states (and hence anti-crossing features) in the InSe monolayer spectrum. We find no evidence for replica bands due to a Moiré superlattice potential. Nonetheless, the measured µARPES spectrum features a pronounced anti-crossing anomaly near the edge of the valence band, as highlighted by a black box in Figure 1a, showing the photoemitted intensity in an energy-momentum slice taken along the **Γ** to **K$_{Gr}$** direction. This "ghost" anti-crossing occurs due to interlayer umklapp hybridization where resonance conditions are achieved by the band crossing between graphene and InSe dispersions in the second Brillouin zone of InSe, also present in the measured spectra of graphene bands (see the purple box in Figure 1a).

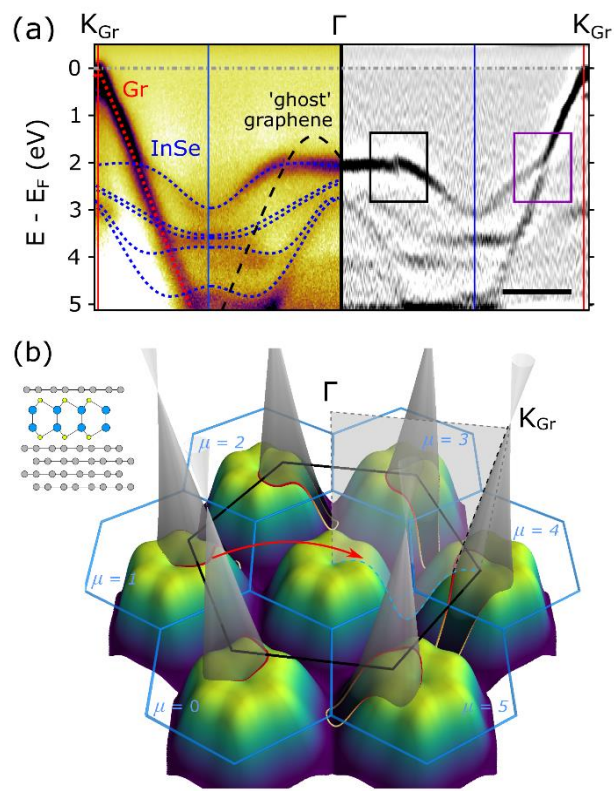

Figure 1: Anti-crossings observed in the valence band structure of graphene/1L InSe heterostructure. (a) left: µARPES energy-momentum slice along the **Γ- K$_{Gr}$** direction, passing through the InSe Brillouin zone boundary (blue vertical line) (**K$_{Gr}$** denotes the corner of graphene's Brillouin zone). Overlaid on the left are theoretical predictions for the isolated InSe bands (blue dashed) and graphene $\pi$ band (red dashed). The black dashed line indicates the position of the graphene $\pi$ band when folded into the 1$^{st}$ Brillouin zone of InSe (note that the folded graphene $\pi$ band does not originate near the Dirac point and so is at higher binding energy). Right: mirrored, the double-differential of the same spectra shown on the left. InSe is at a twist angle of 22.3 ± 0.6° with respect to graphene. An anti-crossing is highlighted by the purple box on the right and a ghost anti-crossing by the black box. Scale bar, 0.5 Å$^{-1}$. (b) 3D

schematic of the uppermost valence band of InSe and graphene $\pi$ band; red/yellow lines highlight the position of overlap. Blue and black hexagons represent InSe and graphene Brillouin zones respectively. The grey plane marks the slice of energy-momentum space covered by the μARPES data in (a). The red arrow indicates folding of an anti-crossing in the second Brillouin zone of InSe to a ghost anti-crossing in the first Brillouin zone. Inset: atomic schematic cross-section of the graphene/1L InSe/graphite 2D heterostructure.

The 2DHSs studied in this work were assembled by dry transfer in an inert environment, where exfoliated crystals of InSe and GaSe were deposited on thick graphite and encapsulated with monolayer graphene (see Methods and schematic inset in Figure 1b). This method allows for ARPES probing of buried layers through graphene (as graphene's ARPES spectrum is already well known [17,18]) while allowing for surface charge dissipation into a conductive substrate (platinum-coated n-Si wafer) [19]. Several samples were fabricated using different thicknesses of InSe and GaSe crystals, and different twist angles with respect to the graphene lattice.

The ghost anti-crossings, and their origin, are more apparent when looking across reciprocal space. The photoemission intensity at a constant energy near the top of the InSe upper valence band (UVB), in a region around $\mathbf{\Gamma}$, is shown in Figure 2a and reveals the twisted lines in reciprocal space at which the ghost anti-crossings occur. The measured data (black dashed rectangle) have been averaged and rotated to form the complete image, as described in Supplementary Material, Section 1. Low energy electron diffraction from a sub-micrometre spot (μ-LEED), taken at the same position on the sample as the μARPES measurements (see low energy electron microscopy image in Supplementary Material, Section 2), gives diffraction peaks from both the graphene and InSe layers. In-plane, InSe has a hexagonal lattice with lattice parameter, $a_{InSe}$ = 4.00 Å [20], 60% larger than that of graphene, $a_{gr}$ = 2.46 Å. When stacked with a twist angle θ between the layers, they form an incommensurate structure where the Brillouin zone corners in the graphene layer, $\mathbf{K_{Gr}}$, lie in the second Brillouin zone of the InSe layer. By identifying the LEED peak positions for both materials, we find the twist angle between their crystalline lattices θ = 22.3 ± 0.6°. In Figure 2d we plot a contour map of the InSe UVB energy in its first and second Brillouin zone, with the first Brillouin zone of graphene overlaid (black hexagon). As shown in the 3-dimensional band schematic, Figure 1b, in monolayers of InSe the UVB dispersion takes the shape of an inverted 'Mexican hat' [21] around the zone centre, $\Gamma$, with the valence band maximum (VBM) close to but not at $\Gamma$, and the band disperses to a minimum at the zone corner $K_{InSe}$. By contrast, the upper valence band in graphene forms the characteristic Dirac cones, meeting the conduction band at the six Dirac points at the zone corners, $K_{Gr}$. Band anti-crossings occur where the graphene and InSe bands would have been coincident. Their position is shown on the contour map by red lines, drawn using an interpolation formula [22], which map out distorted-triangular closed curves around the Dirac cones, in the second Brillouin zone of InSe. Umklapp scattering by an InSe reciprocal lattice vector, $G_{InSe}$, replicates these anti-crossings in the first Brillouin zone of InSe.

This emphasizes the difference between this interlayer Umklapp process and the Moiré phenomena previously reported in systems such as twisted bilayer graphene where a Moiré superlattice potential creates replica bands shifted by the Moiré wave vector $K_M = G_{InSe} - G_{graphene}$ and interaction between the primary and replica bands creates flat bands, as previously observed by μARPES [11,12]. Here, no replica bands are apparent, suggesting negligible Moiré superlattice potential, and the ghost anti-crossings are found by mapping the band anti-crossings by $G_{InSe}$ not $K_M$, similar to previous ARPES measurements of incommensurate twisted bilayer graphene [23]. There are also similarities to the back-folding of bands by charge density waves (CDWs) and spin density waves (SDWs) [24–26], where ARPES has revealed clear anti-crossings between the primary and replica bands even when the replica bands themselves are weak. Here, however, rather than stemming from a new ordered phase within the material, the observed ghost anti-crossings can be explained by interlayer

Umklapp scattering. However, to reproduce the experimentally measured pattern, the angular dependence of the interlayer hybridization must also be considered.

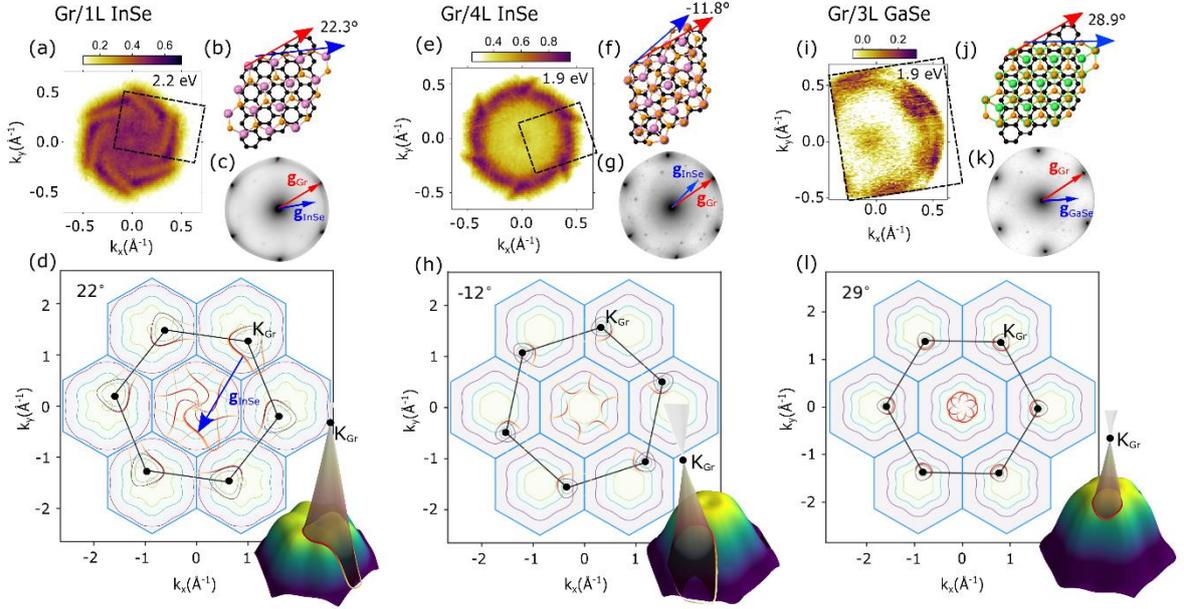

Figure 2: Controlling the position and shape of anti-crossings with twist angle. μARPES constant energy map for (a) graphene/ 1L InSe, (e) graphene/ 4L InSe and (i) graphene/ 3L GaSe at the binding energy stated in the top right corner (averaged over ± 0.05 eV from the stated binding energy). The dashed black rectangles in (a) and (e) are the collected μARPES data, details of the data processing are given in the Supplementary Material Section 1. Colour scales show the normalised photoemission intensity. Relative orientations of the graphene with respect to InSe and GaSe crystals are illustrated in atomic structure schematics (b), (f) and (j) and confirmed by μLEED measurements shown in (c), (g) and (k) respectively. Contour plots of the post-transition metal chalcogenide band structure of each heterostructure are given in (d, h, l). The red lines demonstrate the formation of anti-crossings from the band overlaps: the thickness of the red line indicates the size of the gap, and the shade of red indicates the energy at which the gap occurs, as illustrated in the 3D schematics, bottom right.

To do this, we employ a method developed [17,27] for the description of ARPES intensity maps of graphene [28] and tunnelling between 2D crystals [9,29]. It uses a plane wave decomposition of Bloch states of electrons in the graphene bands, and in the UVB of InSe, then describes the variation of the interlayer hybridisation parameters across the relevant part of the Brillouin zone by projecting the plane wave components with the coinciding wave vectors. The states involved in the hybridisation are related by the umklapp condition, $\mathbf{q} = \xi \mathbf{K}_n - \mathbf{G}_{InSe} + \mathbf{p}$ (where $\xi = +$ or $-$ from the two inequivalent valleys in graphene and $n = 0, 1$ and 2 indexes the three equivalent valleys, $\mathbf{p} = (p_x, kp_y)$ is the valley momentum of graphene, and $\mathbf{q}$ is the momentum of the InSe counted from the Brillouin zone centre with $\mathbf{q} \ll \mathbf{G}_{InSe}$). Along the six lines that satisfy this condition (Figure 2d), counted by integer $\mu = 0$ to 5 counter-clockwise, hybridisation between the graphene-InSe bands leads to the anti-crossing of their dispersions, splitting them apart by

$$\delta\varepsilon \propto \left|\sin(\varphi_{\vec{p}} - \frac{\pi}{3}\mu)\right|, \qquad (1)$$

where $\varphi_{\mathbf{p}} = \arctan(p_y/p_x)$ describes the direction of valley momentum. The angular dependence highlighted in equation (1) is the result of sublattice composition-n (AB) of electron states in graphene (for details see Methods), which has been seen in twist-controlled electron tunnelling in graphene/hBN/graphene heterostructures [9]. On the plot in Figure 2d, this variation of the interlayer hybridisation factor, $\delta\varepsilon$, is shown by the thickness of the red lines. Taking this variation into account, the resultant momentum-space images of the anti-crossings qualitatively reproduce the experimental behaviour observed in Figure 2a.

The pattern of anti-crossings is dependent on θ (see Supplementary Material, Section 3), and can be observed in other 2D heterostructures. Figure 2e shows the mini-gap pattern apparent in a constant energy slice taken on graphene-capped 4L InSe, with θ = 11.8 ± 0.1° determined by µ-LEED. Again, a contour plot showing the expected position of the ghost anti-crossings, Figure 2h, calculated by umklapp scattering from the overlap contours in the second Brillouin zone of InSe, qualitatively reproduces the experimental measurement. We observe the same effect in a constant energy slice from a 2D heterostructure of graphene capped 3L GaSe (Figure 2i), where µ-LEED was used to determine θ = 28.9 ± 0.7°. The minigap pattern observed in the experimental measurement again agrees with the expected pattern, Figure 2l. The "ghost" anticrossing is observed in the energy-momentum spectra shown in Supplementary Material, Section 4.

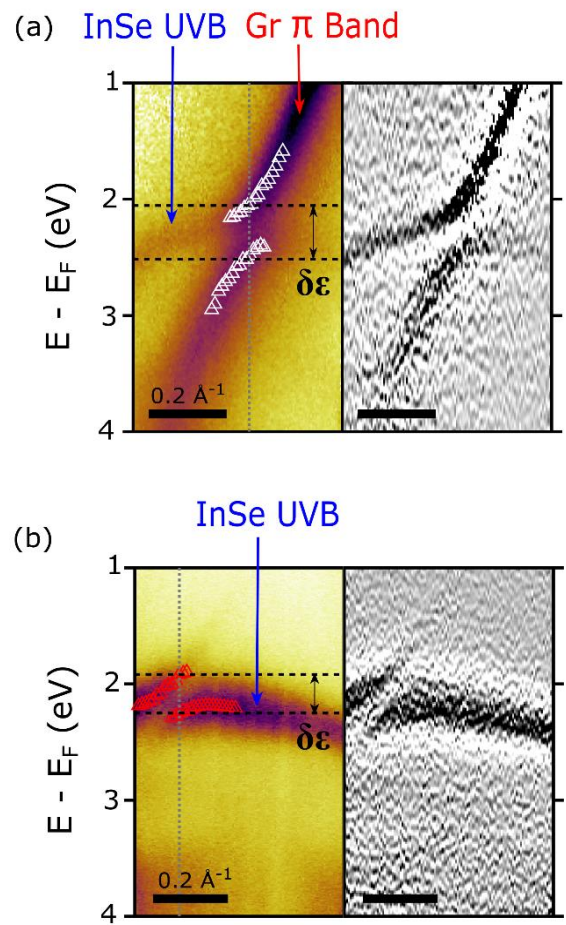

Figure 3: Magnitude of anti-crossing. (a) μARPES measurement of the anti-crossing created between the graphene $\pi$ band and 2L InSe upper valence band, with the double-differential (with respect to energy) of the same spectra on the right. Overlaid white triangles show the fit of band positions. Energy distribution curves with fits can be seen in Supplementary Material, Section 4. (b) μARPES measurement of the ghost anti-crossing in the upper valence band of 4L InSe, with the double-differential of the same spectra on the right. Overlaid red triangles show the fit of band positions. These spectra were from the same heterostructure, in different positions on the InSe flake (see Supplementary Material, Section 4).

The interlayer hybridization factor, $\delta\varepsilon$, can be determined from fitting the band dispersions where the InSe UVB meets the graphene $\pi$ band. An ARPES energy-momentum slice of the anti-crossing, from the same sample as Figure 2e, is shown in Figure 3. The band mixing and anti-crossing gap are clear in the data twice-differentiated with respect to energy, the right spectra of Figure 3. Band positions were found by fitting energy distribution curves (EDC) around the anti-crossing, and from these band positions the interlayer hybridization factor was found, $\delta\varepsilon = 0.45 \pm 0.02$ eV. The dispersion and magnitude of this anti-crossing are consistent with the corresponding ghost anti-crossing, as shown in Figure 3b from a region of graphene covering 4L InSe. This clear ghost anti-crossing in 4L InSe demonstrates that the effect is not limited to monolayer materials. Note that Figure 3a and 3b were acquired from different areas of the same graphene on InSe heterostructure with the same orientations of graphene and InSe flakes but differing InSe thickness (see Supplementary Material, Section 5).

Ghost anti-crossings are also apparent in *ab initio* predictions of the band structure of the composite graphene / InSe stack, further confirming that they are an inherent feature of the electronic structure. Using linear scaling DFT [30,31], we studied graphene on monolayer InSe at 23°, corresponding to a supercell of 698 atoms (see Methods and Supplementary Material, Section 6). Using previously reported tools to project the electronic structure into the primitive cells of each layer [32], simulated spectra of the valence bands of the composite structure were constructed. A momentum slice plotted along **Γ** to **K**$_\text{Gr}$ (Figure 4a) shows mini-gaps at anti-crossings of InSe valence bands with the graphene $\pi$ band, as well as anti-crossings with interlayer umklapp scattered $\pi$ bands. These are not present in the electronic structures of the separated layers. A simulated constant energy slice (Figure 4b) shows the same characteristic vortex as observed experimentally and as predicted by the analytical model above. Changes in the measured and simulated pattern with binding energy are compared in Supplementary Material, Section 7.

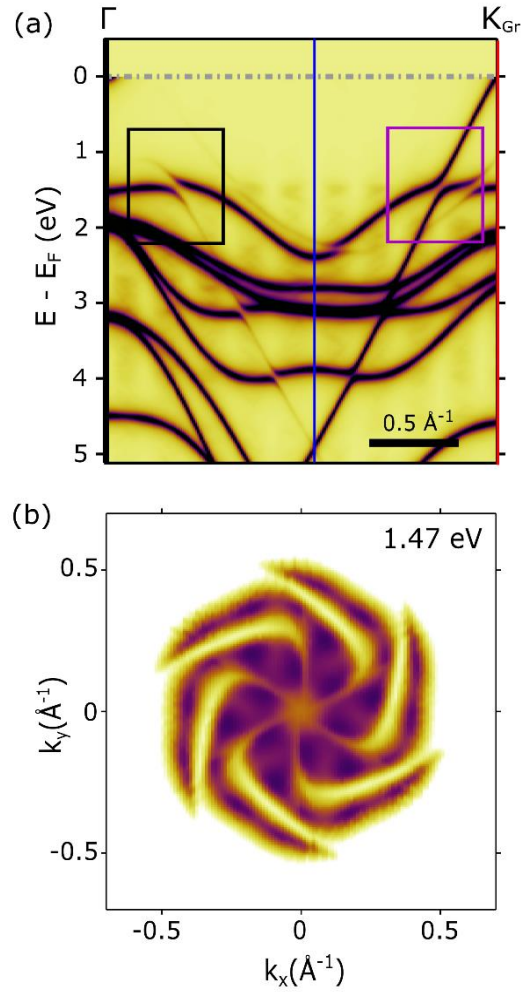

Figure 4: Ab-initio calculation of graphene/ 1L InSe bandstructure. (a) DFT calculated energy-momentum spectrum along the **Γ- K$_{Gr}$** direction, passing through the InSe Brillouin zone boundary (blue vertical line). The twist angle between graphene and InSe is close to 23°, similar to the heterostructure measured in Figure 1a. An anti-crossing is highlighted by the purple box on the right, and a ghost anti-crossing by the black box. Comparison to μARPES spectra can be seen in Supplementary Material, Section 6. (b) Constant energy map at close to the VBM (binding energy stated in the top right corner).

# Conclusion

In summary, our data present evidence for interlayer umklapp scattering in 2D heterostructures. The ghost anti-crossings created near the valence band maximum of monolayer InSe demonstrate points of strong coupling with the adjacent graphene layer, with their position controlled by the relative orientation between the layers. Further control could be gained through changing band-alignments, using chemical doping or a perpendicular electric field [33]. By selecting suitable 2DMs pairs, it should thus be possible to engineer strong mixing of their electronic states at or near the band edges of many 2D semiconductors, or near the Fermi-level of metals and semi-metals, giving a new tool for band structure engineering. This interlayer umklapp scattering should not be limited to band hybridisation, and we expect it also to manifest in further novel electron, phonon and photon interactions in 2D heterostructures.

# Methods

***Sample Fabrication.*** Bulk rhombohedral γ-InSe crystals, purchased from 2D Semiconductors and grown using the vertical Bridgman method, were mechanically exfoliated down to thin (1L to 10L) crystals on a silicon oxide substrate. Using the PMMA dry peel transfer technique [34], monolayer graphene was used to pick-up and stamp InSe crystals onto either graphite or hBN, each of which was laterally large (>50 μm), thin (<50 nm) and positioned on a (3 nm) Ti/ (20 nm) Pt - coated highly n-doped silicon wafer. Both heterostructure samples were annealed to 150° for 1 hour in order to remove impurities via the self-cleaning mechanism[6]. All that is described above took place within an Ar glove box to prevent sample degradation. The same samples were used for both µARPES and µLEED measurements.

**µARPES.** µARPES spectra were acquired from the Spectromicroscopy beamline of the Elettra light source[35]. A low energy (27 eV), linearly polarised photon beam was focused onto the sample surface using Schwarzschild objectives. The beam had a submicrometre spot size (~ 600 nm) and was at an incident angle of 45° to the sample surface (linearly polarised at 45° to the sample surface). To perform ARPES, photoemitted electrons were collected by an internal moveable hemispherical electron analyser and 2D detector with an energy and momentum resolution of ~50 meV and ~0.03 Å$^{-1}$. Before analysis, samples were annealed for > 6 hours at up to around 625 K in ultra-high vacuum. The correct position on the sample was found by comparing an optical image of the specimen to scanning photoemission microscopy (SPEM) images acquired on the beamline before ARPES measurements. Energy-momentum slices along the high symmetry directions of the Brillouin zones were acquired by measuring a series of closely spaced detector slices and interpolating the spectra. The constant energy maps, $I(k_x, k_y)$, around $\Gamma$ were extracted from 3-dimensional energy-momentum maps, $I(E, k_x, k_y)$ and averaged over an energy range of 0.04 eV. The sample temperature during measurement was ~100 K.

**µLEED.** µLEED patterns were acquired on the low energy electron microscope (LEEM) at the Nanospectromicroscopy beamline of the Elettra light source [36]. A well-collimated beam of low energy electrons from a LaB$_6$ electron-gun was focused on to the sample, the electron energy being set by applying a voltage bias to the sample stage. An e-beam footprint on the sample of only 500 nm was obtained by inserting an illumination limiting aperture in the microscope optical path. Magnified images of the diffraction pattern produced by elastically backscattered electrons were acquired using a 2D detector and CCD. The diffraction pattern shown in Figure 2c is an average of multiple diffraction patterns collected over an incident electron energy range of 30-60 eV in steps of 2 eV. The diffraction pattern shown in Figure 2g is an average of multiple diffraction patterns collected over an incident electron energy range of 27-60 eV in steps of 1 eV. The diffraction pattern shown in Figure 2k was collected with an incident electron energy of 55 eV.

*Modelling of sublattice effects on resonant hybridization between graphene and InSe.*

We used a plane wave decomposition of Bloch states of electrons in the graphene bands, and in the UVB of InSe, then projected the plane wave components with the coinciding wave vectors. For graphene, the relevant parts of the spectrum appear in the vicinity of Dirac points near K ($\xi = +$) and K' ($\xi = -$) valleys, where the plane wave decomposition, that involves the set of smallest wave vectors $\xi \vec{K}_n + \vec{p}$, related by the reciprocal lattice vectors for graphene, reads

$$\psi_\xi \approx N(|z|) \sum_{K_{Gr}} \left[1 + s\xi e^{i\xi(\varphi_{\vec{p}} - \frac{2\pi}{3}n)}\right] e^{i(\xi \vec{K}_{Gr} + \vec{p}) \cdot \vec{r}}.$$

Here, $\varphi_{\vec{p}} = \arctan(p_y/p_x)$ describes the direction of valley momentum $\vec{p} = (p_x, p_y)$ of electrons in graphene, $s = +1$ for conduction band and $s = -1$ for valence band branch of dispersion ($s = -1$ is the one relevant for resonant mixing with InSe valence band states), $\vec{r} = (x, y)$, and $N(|z|)$ takes into account the decay of the 2D plane wave amplitude away from the crystal. The factor $\left[1 + s\xi e^{i\xi(\varphi_{\vec{p}} - \frac{2\pi}{3}n)}\right]$ accounts for the interference of the contributions to the 'vacuum' planes coming from $P_z$ orbitals of carbons on two (A&B) sublattices of honeycomb graphene[17]. For InSe, we concentrate on the UVB dispersion in the vicinity of its top near the BZ centre, where the electron states come mostly from S and $P_z$ orbitals of chalcogen atoms [37]. This prescribes the plane wave decomposition for the top valence band state in monolayer InSe,

$$\Psi \approx f(|z|) e^{i\vec{q} \cdot \vec{r}} + g(|z|) \sum_\mu e^{i(\vec{G}_{InSe} + \vec{q}) \cdot \vec{r}},$$

where $\vec{G}_{InSe}$ is the first star of InSe reciprocal lattice vectors and $\vec{q}$ is counted from the BZ centre ($q \ll G_{InSe}$). After projecting the plane waves in the two crystals, we find that the states involved in the hybridisation are related by the Umklapp condition, $\vec{q} = \xi \vec{K}_n - \vec{G}_{InSe} + \vec{p}$, and the variation of their interlayer coupling across the BZ is described by $h(\vec{q}) \propto \left[1 + s\xi e^{i\xi(\varphi_{\vec{p}} - \frac{2\pi}{3}n)}\right]$. Due to the decay of the electronic wave functions away from each crystal, which is even faster for the $\vec{G}_{InSe} + \vec{q}$ plane wave components ($g(|z|)$) than for the $\vec{q}$ component ($f(|z|), i.e. g(|z|) \ll f(|z|)$ at distances $|z|$ longer than the Bohr radius), the hybridisation of graphene and InSe bands would be negligibly weak, unless it satisfies the resonant condition, $\varepsilon_{InSe}(\vec{q}) = \varepsilon_{gr}(\vec{p})$. Along the six lines identified by such crossings, counted by integer $\mu = 0$ to 5 counter-clockwise, graphene-InSe bands hybridisation leads to the anti-crossing of their dispersions, splitting them apart by $\delta\varepsilon = 2|h| \propto \left|\sin(\varphi_{\vec{p}} - \frac{\pi}{3}\mu)\right|$, leading to Eq.1 in the main text. Values of $\vec{G}_{InSe}$= 1.73 Å$^{-1}$ and $\vec{G}_{GaSe}$ = 1.85 Å$^{-1}$ were used, consistent with literature values for these materials [38].

*Ab initio calculations.* Linear-scaling DFT (LS-DFT) calculations in the Projector Augmented Wave formalism [30,39] were used to model the InSe/Gr heterostructure, using the ONETEP code [31]. Further details are given in Supplementary Material, Section 3.

# Acknowledgements

This work was supported by EPSRC (grants EP/N509565/1, EP/P01139X/1, EP/N010345/1, and EP/L01548X/1 along with the CDT Graphene-NOWNANO and the EPSRC Doctoral Prize Fellowship), the European Graphene Flagship Project, ERC Synergy Grant Hetero2D, the ARCHER National UK Supercomputer RAP Project e547, Royal Society URF, and Lloyd Register Foundation Nanotechnology grant. We acknowledge the use of Athena at HPC Midlands+, which was funded by the EPSRC through Grant No. EP/P020232/1, as part of the HPC Midlands+ consortium, and computing resources provided by the Scientific Computing Research Technology Platform of the University of Warwick. The research leading to this result has been supported by the project CALIPSOplus under Grant Agreement 730872 from the EU Framework Programme for Research and Innovation HORIZON 2020. The authors declare no competing interests.

# Supplementary material for 'Ghost anti-crossings caused by interlayer umklapp hybridisation of bands in 2D heterostructures'

*Abigail J. Graham[1], Johanna Zultak[2,3], Matthew J. Hamer[2,3], Viktor Zolyomi[2,3], Samuel Magorrian[2,3], Alexei Barinov[4], Viktor Kandyba[4], Alessio Giampietri[4], Andrea Locatelli[4], Francesca Genuzio[4], Natalie Teutsch[1], Cuauhtémoc Salazar[1], Nicholas D. M. Hine[1], Vladimir I. Fal'ko\*[2,3,5], Roman V. Gorbachev\*[2,3,5], Neil R. Wilson\*[1]*

[1]Department of Physics, University of Warwick, Coventry, CV4 7AL, U.K.

[2]National Graphene Institute, University of Manchester, Oxford Road, Manchester, M13 9PL, U.K.

[3]School of Physics and Astronomy, University of Manchester, Oxford Road, Manchester, M13 9PL, U.K.

[4]Elettra – Sincrotrone Trieste, S.C.p.A, Basovizza (TS), 34149, Italy

[5]Henry Royce Institute, Oxford Road, Manchester, M13 9PL, U.K.

## Contents

# Section 1: Constant energy map image processing

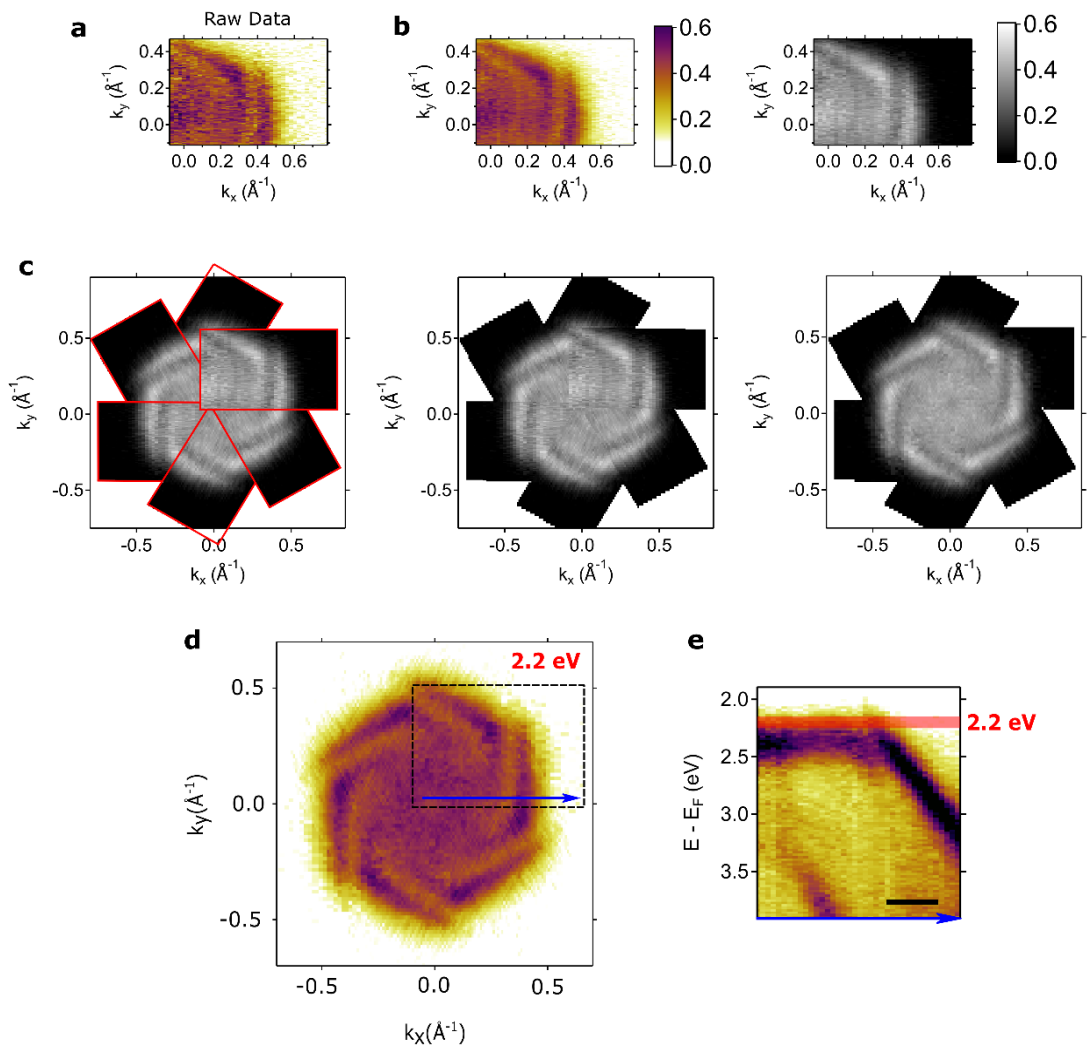

**Figure S1: Constant energy map image processing**. The μARPES raw data is shown in (a) for a single binding energy of 2.20 eV. (b) The raw data averaged over 0.1 eV (2.15-2.25 eV) and presented in two different colour schemes. The construction of the symmetrized spectrum, is shown in (c): the data in (b) is rotated about (0, 0) for multiples of 60°, and these rotated spectra are stacked on top of one another as displayed in the left (with red boxes to show the stacking order) and middle images (boxes removed); the overlapping data points are then averaged to give the right image. (d) The same image as the right of (c) in the colour scheme used in Figure 2a in the main text with black dashed rectangle showing the area of the original data. The blue arrow shows the position of the energy-momentum slice in (e). In (e), the red line shows the energy range averaged (0.1 eV) to produce the constant energy map. Scale bar, 0.2 Å$^{-1}$. The same processing was also used to create Figure 2e in the main text and the constant energy maps in Figure S3 and Video S2. Colour scales show the normalized photoemission intensity (arbitrary units).

## Section 2: Microscopy of graphene on 1L InSe 2D from Fig 1 heterostructure.

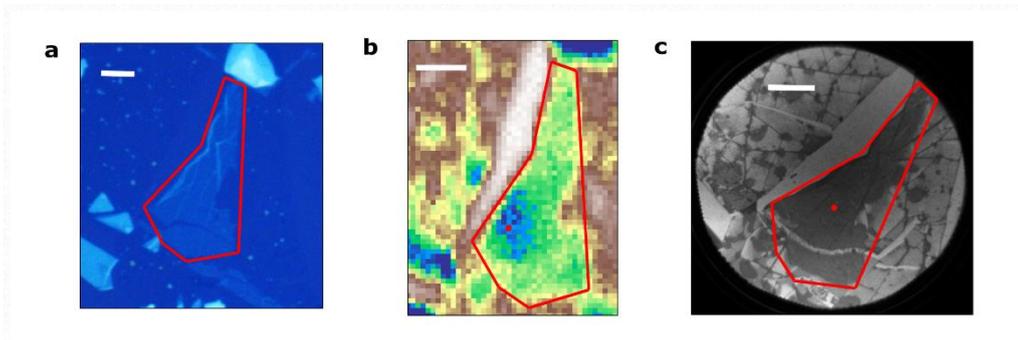

**Figure S2: Microscopy of the heterostructure.** Optical (a), SPEM (b) and LEEM (c) images. 1L InSe flake on hBN, fully encapsulated by graphene. SPEM image taken with centre energy 21 eV. LEEM image taken with a start voltage of 13.6 eV. 1L InSe flake highlighted by red outline. Red spot indicates position of ARPES measurement in (b) and μLEED measurement in (c). All scale bars, 5 μm.

## Section 3: Animation of the dependence of ghost anti-crossings on twist angle

Video S1: Dependence of ghost anti-crossings on twist angle. 3D schematic and contour plot of 1L InSe upper valance band and graphene π band for different twist angles showing how the shape and position of the hybridization gap changes.

# Section 4: Microscopy of graphene / GaSe 2DHS from Fig. 2

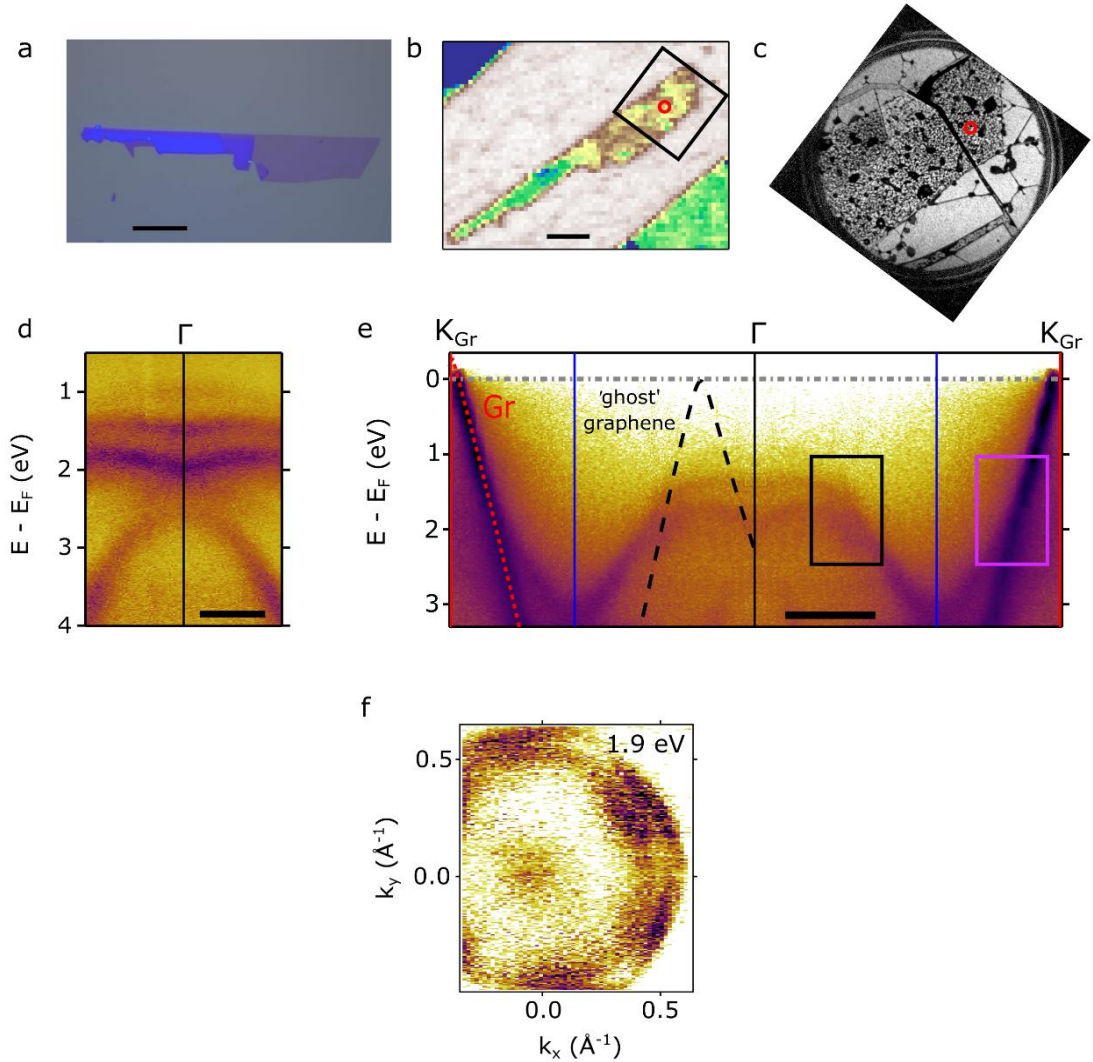

**Figure S3 : Microscopy of graphene / GaSe 2DHS from Fig 2.** (a) Optical microscopy image of the GaSe flake, different optical contrast corresponds to differing numbers of layers of GaSe. Scale bar, 10 μm. (b) Scanning photoemission microscopy image of the GaSe flake. The red circle indicates the position of the measurement in Figure 2i. The black box shows the region shown in (c). Scale bar, 10 μm (c) LEEM image of the heterostructure. The red circle indicates position of measurement in Figure 2k. (d) μARPES energy-momentum slice around Γ, showing that the GaSe flake is 3 layers thick. Scale bar, 0.2 Å$^{-1}$. (e) μARPES energy-momentum slice along the **Γ- K$_{Gr}$** direction, passing through the GaSe Brillouin zone boundary (blue vertical line) (**K$_{Gr}$** denotes the corner of graphene's Brillouin zone). Overlaid on the left are theoretical predictions for the graphene π band (red dashed). The black dashed line indicates the position of the graphene π band when folded into the 1$^{st}$ Brillouin zone of GaSe. GaSe is at a twist angle of 28.9 ± 0.7° with respect to graphene. An anti-crossing is highlighted by the purple box on the right and a ghost anti-crossing by the black box. The anticrossing in the purple box is not visible due to the high intensity of the graphene π band, but within the black box we see a clear deviation of the GaSe valence band from the expected continuous dispersion in a position corresponding to the folded graphene π band. Scale bar, 0.5 Å$^{-1}$. (f) Raw data averaged over 0.1 eV used to create Figure 2i.

# Section 5: Microscopy of graphene / InSe 2DHS from Fig 3

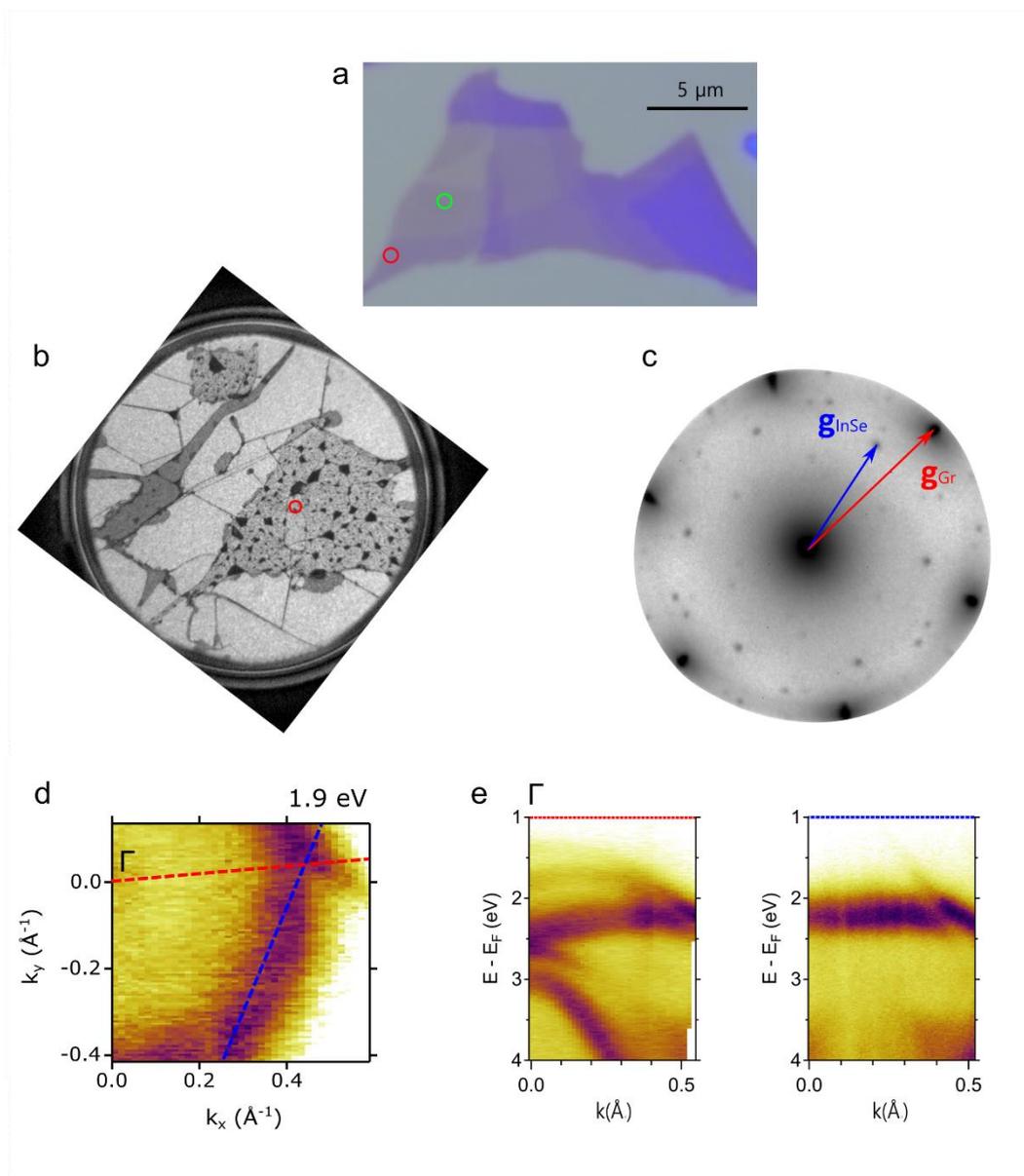

**Figure S4: Microscopy of graphene / InSe 2DHS from Fig 3**. (a) Optical microscopy image of the InSe flake, different optical contrast corresponds to differing numbers of layers of InSe. The green (red) circle indicates the position of the measurement in Figure 3a (Figure 3b). (b) LEEM image of the heterostructure and (c) LEED pattern from the region marked by the red circle in (b). (d) Raw data averaged over 0.1 eV used to create constant energy map in Figure 2e. Spectra acquired from the 4L region at the position marked by the red circle in (a). (e) Energy-momentum slices extracted along (left) the red dashed line and (right) the blue dashed line in (d).

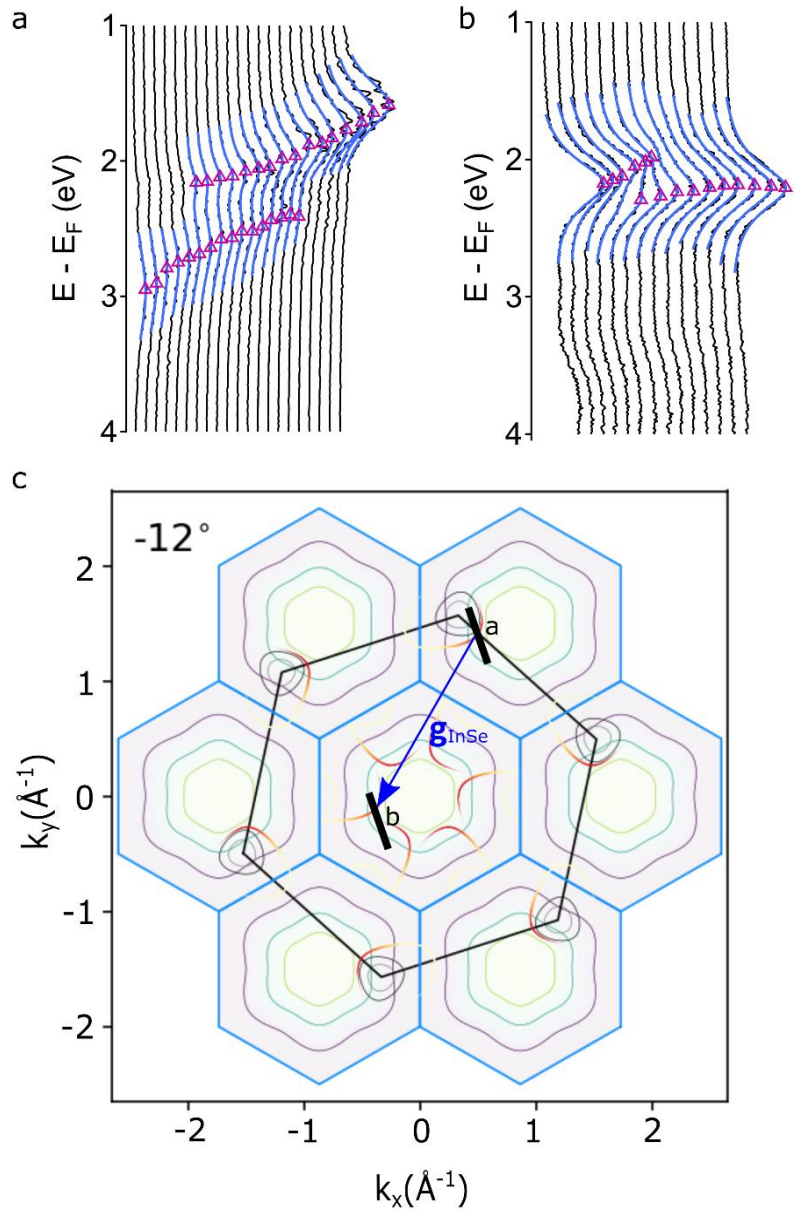

**Figure S5: Schematic of the positions of the energy-momentum slices shown in Fig 3 and their energy distribution curves**. a) and b) Energy distributions curves from Figure 3a and 3b. Overlaid in blue are the Lorentzian peak fits, with the peak positions shown by purple triangles. Note that for clarity, alternate data lines are shown giving fewer peak positions than shown in the main text. c) Contour plots of the band structure for the InSe / graphene heterostructure of Figure 3. The blue hexagons mark the Brillouin zones of InSe and the black hexagon the first Brillouin zone of graphene. The solid black lines labelled a and b correspond to the positions in momentum space of the energy-momentum spectra shown in Figure 3a and 3b respectively.

# Section 6: Band structure calculations.

The valence band structure for isolated 1L InSe, overlaid in Figure 1a, was calculated using first principles density functional theory (DFT) as implemented in the VASP code [1]. The in-plane lattice parameter was taken to be 4.00 Å and the interlayer distance was taken to be 8.32 Å, found from experiments [2]. The atomic positions were relaxed until forces on the atoms were less than 0.005 eV/Å. The plane wave cut-off energy was 600 eV and the Brillouin zone was sampled with a 24 × 24 × 1 grid. Spin-orbit coupling was taken into account for the calculations. The local density approximation (LDA) exchange-correlation functional was used, with a scissor correction to correct for the underestimated band gap. The calculated bands were shifted by 1.25 eV to align to the valence band edge in the ARPES spectrum. The valence band structure of the graphene π band overlaid in Figure 1a was calculated using the tight-binding model [3], with parameters $t$ and $t'$ equal to -3.2 eV and 0.0 eV respectively.

The contour plots and 3D schematics presenting the valence band energy surfaces for 1L InSe (Figure 1b and Figure 2d), 4L InSe (Figure 2h) and 3L GaSe (Figure 2l) were obtained using the tight-binding model of ref. [4] originally developed for few-layer InSe. The parameters re-fitted to scissor-corrected DFT data for few-layer GaSe [5] are shown in the table below.

| | | | | | |
|---|---|---|---|---|---|
| $\varepsilon_{M_s}$ | -0.879 | $T^{(3)}_{ss}$ | 0.824 | $t^{(XX)}_{ss}$ | -0.568 |
| $\varepsilon_{M_{px}} = \varepsilon_{M_{py}}$ | 3.779 | $T^{(3)}_{Ms-Xp}$ | 0.217 | $t^{(XX)}_{sp}$ | -0.549 |
| $\varepsilon_{M_{pz}}$ | 7.871 | $T^{(3)}_{Mp-Xs}$ | -0.379 | $t^{(XX)}_{\pi}$ | -0.120 |
| $\varepsilon_{X_s}$ | -8.489 | $T^{(3)}_{\pi}$ | 0.002 | $t^{(XX)}_{\sigma}$ | -0.751 |
| $\varepsilon_{X_{px}} = \varepsilon_{X_{py}}$ | -1.849 | $T^{(3)}_{\sigma}$ | -0.607 | $t'^{(MX)}_{ss}$ | 0.416 |
| $\varepsilon_{X_{pz}}$ | -1.388 | $T'^{(1)}_{ss}$ | -1.837 | $t'^{(MX)}_{Ms-Xp}$ | 0.195 |
| $T^{(1)}_{ss}$ | -0.058 | $T'^{(1)}_{sp}$ | -4.996 | $t'^{(MX)}_{Mp-Xs}$ | -0.518 |
| $T^{(1)}_{Ms-Xp}$ | 3.241 | $T'^{(1)}_{\pi}$ | -0.819 | $t'^{(MX)}_{\pi}$ | -0.003 |
| $T^{(1)}_{Mp-Xs}$ | -2.366 | $T'^{(1)}_{\sigma}$ | -4.384 | $t'^{(MX)}_{\sigma}$ | -0.109 |
| $T^{(1)}_{\pi}$ | 1.203 | $T^{(2)}_{ss}$ | 0.441 | $t'^{(XM)}_{ss}$ | -0.464 |
| $T^{(1)}_{\sigma}$ | 2.146 | $T^{(2)}_{Ms-Xp}$ | -0.509 | $t'^{(XM)}_{Ms-Xp}$ | -0.168 |
| $T^{(2X)}_{ss}$ | -1.256 | $T^{(2)}_{Mp-Xs}$ | 0.397 | $t'^{(MX)}_{Mp-Xs}$ | -0.151 |
| $T^{(2X)}_{sp}$ | -1.045 | $T^{(2)}_{\pi}$ | 0.138 | $t'^{(MX)}_{\pi}$ | -0.286 |
| $T^{(2X)}_{\pi}$ | -0.251 | $T^{(2)}_{\sigma}$ | 0.246 | $t'^{(MX)}_{\sigma}$ | 0.382 |
| $T^{(2X)}_{\sigma}$ | -0.724 | $T'^{(3)}_{ss}$ | -0.330 | | |
| | | $T'^{(3)}_{sp}$ | -0.075 | | |
| | | $T'^{(3)}_{\pi}$ | 0.022 | | |
| | | $T'^{(3)}_{\sigma}$ | 0.052 | | |

Linear-scaling DFT (LS-DFT) calculations in the Projector Augmented Wave formalism [6,7] were used to model the InSe/Gr heterostructure, using the ONETEP code [8]. The optB88 van der Waals density functional [9] was used as it has been shown to give reliable in-plane and interlayer distances for this class of materials [10]. We used PAW datasets from the GBRV library [11], and applied Ensemble DFT [12], using an NGWF radii of 12 $a_0$ and a psinc grid of 1200 eV, which are sufficient for well-converged band structures for these materials. We modelled the heterostructure at the same alignment angle, 23°, as in the experimental system in Figure 1 and Figure 2a. A specific pair of choices of supercell then gives a very low-strain commensurate structure of 698 atoms (300 InSe, 398 C) where no strain component exceeds 0.75%. Geometry optimisation of the individual layers, followed by interlayer distance optimisation (resulting in a centre-to-centre distance of 6.19 Å), produced a geometry for the heterostructure where no forces exceed 0.12 eV/Å so further geometry optimisation was not required. The eigenvalues of the supercell calculation were unfolded to the primitive cell

Brillouin zones of the individual layers in a spectral function unfolding calculations as described in ref [10], then combined along a 240-point **Γ** to **K_Gr** path and broadened with a Lorentzian of width 0.05 eV to produce Figure 4a. We used 32 paths of 40 k-points from **Γ** to endpoints along the **K_InSe** - **K'_InSe** line to produce Figure 4b (and DFT constant energy maps in Figure S2).

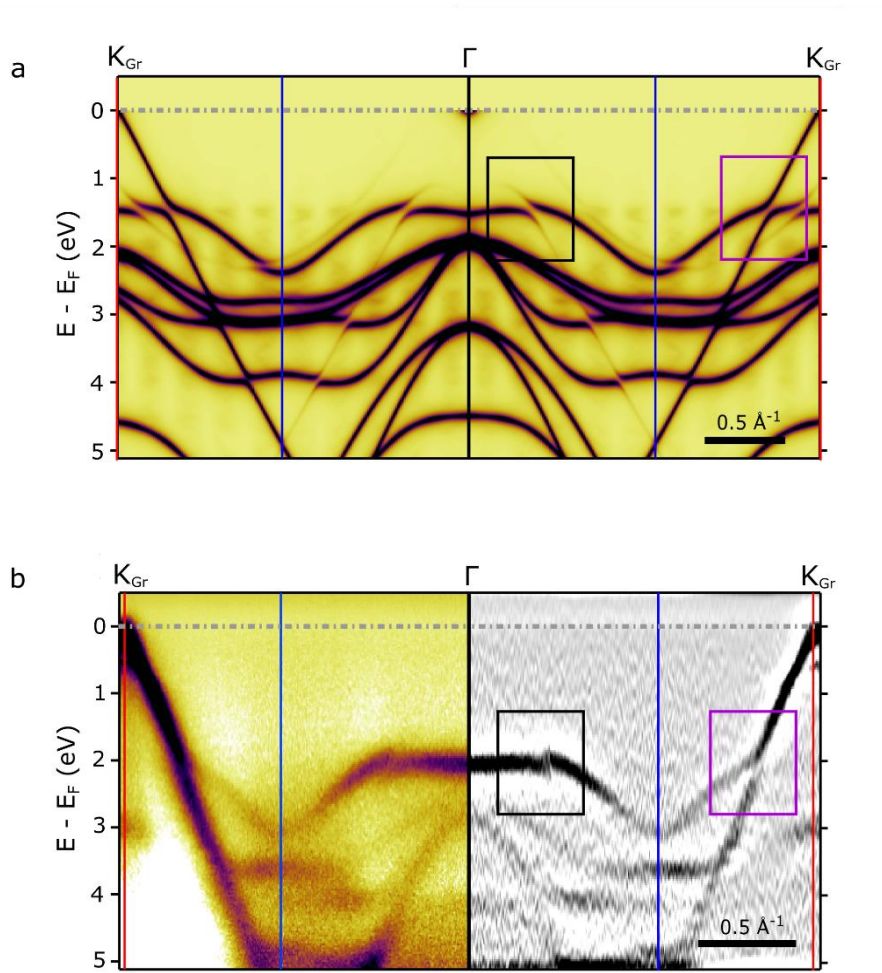

**Figure S6: LS-DFT comparison to µARPES spectra.** a) LS-DFT. Mirrored spectra of Figure 4a. Twist angle of 23° between graphene and InSe. b) µARPES energy-momentum slice from Figure 1a. Right: mirrored, the double-differential of the same spectra shown on the left. InSe is at a twist angle of 22.3 ± 0.6° with respect to graphene. InSe Brillouin zone boundary is shown by the blue line. An anti-crossing is highlighted by the purple box on the right and a ghost anti-crossing by the black box.

# Section 7: Dependence of ghost anti-crossings pattern on binding energy

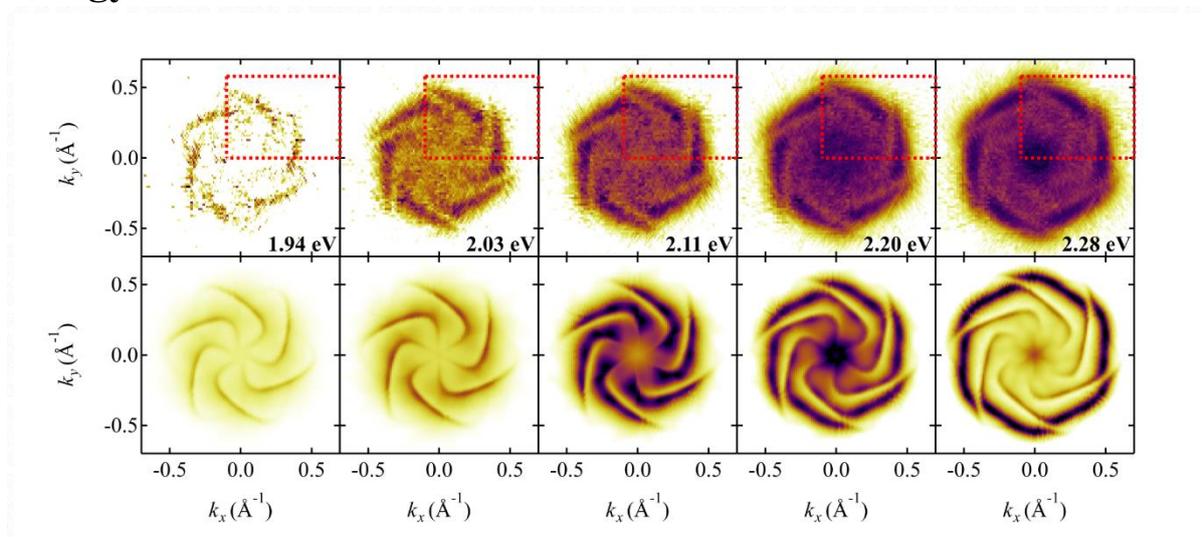

**Figure S7: Shape of avoided crossing dependence on binding energy.** Top panel: μARPES constant energy maps around Γ for graphene on 1L InSe. Lower panel: corresponding DFT calculated constant energy maps around Γ. Binding energy notated in bottom right of μARPES maps. The collected μARPES data is shown within the red dashed box. Due to the overestimated band gap in DFT calculations, the corresponding DFT constant energy maps were found from taking the same difference in energy from the valence band maximum on the InSe in the DFT and μARPES spectra. Also see Video S2 for additional constant energy maps.

**Video S2: Dependence of ghost anti-crossings on binding energy.** Left: μARPES constant energy maps around Γ for graphene on 1L InSe. Right: corresponding DFT calculated constant energy maps around Γ. Binding energy notated in bottom right of μARPES maps. The collected μARPES data is shown within the red dashed box. DFT constant energy maps are at the same energy below the valence band maximum on the InSe in the DFT as the μARPES spectra.